\documentclass [12pt]{article}
\usepackage{hyperref}

\usepackage{graphicx}

\def\beq{\begin{equation}}
\def\eeq{\end{equation}}
\def\bea{\begin{eqnarray}}
\def\eea{\end{eqnarray}}

\begin{document}

\title { Unbound exotic nuclei studied by projectile fragmentation.}

\author { G. Blanchon$^{(a)}$, A. Bonaccorso$^{(a)}$, D. M.  
Brink$^{(b)}$,\\A. Garc\'ia-Camacho $^{(a)}$
 and N. Vinh Mau$^{(c)}$\\
\small $^{(a)}$ Istituto Nazionale di Fisica Nucleare, Sez. di
Pisa, \\
\small and Dipartimento di Fisica, Universit\`a di Pisa,\\
\small Largo Pontecorvo 3, 56127 Pisa, Italy.\\
\small $^{(b)}$ Department of Theoretical Physics, 1 Keble Road, Oxford  
OX1 3NP, U. K.\\
\small $^{(c)}$ Institut de Physique Nucl\'eaire, IN2P3-CNRS, F-91406,  Orsay Cedex, France. }

\maketitle

\begin{abstract}

We present a simple time dependent model for the excitation of a  
nucleon from a bound state to a continuum resonant state in a
neutron-core complex potential which acts as a final state interaction.  
   The
final state is described by an optical model S-matrix so that both  
resonant and non resonant states of any continuum energy can be studied as well
as deeply bound initial states. It is shown that, due to the coupling  
between the initial and final states, the neutron-core free
particle phase shifts are modified, in the exit channel, by an  
additional phase. The effect of the additional phase  on  
the breakup spectra is clarified.  As an example  the population of the low  
energy resonances of   $^{11}$Be and of the unbound
$^{13}$Be  is discussed. Finally, we suggest that the excitation energy spectra of an  
unbound nucleus might reflect the structure of the parent nucleus from
whose fragmentation they are obtained. \end{abstract}

\section{Introduction}

In this paper we will call {\it projectile fragmentation}  the well known elastic breakup (diffraction reaction) of neutron halo nuclei, when the observable studied is the neutron-core relative energy spectrum. This kind of observable has been widely  measured in relation to the Coulomb breakup on heavy target. Recently results on light targets have also been presented  \cite{fuku}. These data enlighten the effect of the neutron final state interaction with the core of origin, while observables like the core energy or momentum distributions enlighten the effect of the neutron final state interaction with the target.

Projectile fragmentation has been used experimentally also with
  two neutron halo projectiles. In this case  it has been suggested that the  reaction might proceed in  
one step (simultaneous emission of the two neutrons) or two steps
(successive emissions) depending on whether the target is heavy and  
therefore Coulomb breakup (core recoil) is the dominant
mechanism or the target is light and then nuclear breakup is the  
dominant mechanism \cite{nig}. The successive emission can be due to different mechanisms. One possibility is that one neutron is ejected because of the interaction with the target, as in the one-neutron fragmentation case, while the other is left behind, for example in a resonance state, which then decays. This second step has been described by the sudden approximation  in Ref.\cite{bhe} under the hypothesis that the first neutron is stripped and that the transparent limit for the second neutron applies.  It corresponds to consider the second neutron emitted at large impact parameters such that the neutron-target interaction can be neglected. The two-step mechanism implies that the two neutrons are not strongly correlated such that the emission can be considered sequential.

However the neutron-target interaction gives rise not only to  stripping but also to elastic breakup and in both cases to first order in the interaction the neutron  ends-up in a plane wave final state \cite{bb1}. It can then re-interact with the core which, for example,  is going to be $^{10}$Be in the case of the one-neutron halo projectile $^{11}$Be, while it will be  $^{12}$Be in the case of the projectile fragmentation of $^{14}$Be, since $^{13}$Be is not bound. While in the case of $^{11}$Be the structure of both its bound and continuum states is well known from other kinds of experiments and  therefore projectile fragmentation experiments are useful to enlighten the reaction mechanism and its possible description, in the case of $^{13}$Be or of other unbound nuclei the interplay between structure and reaction aspects is still  to be clarified.

 Experiments with a  $^{14}$B projectile \cite{jl,Usa} have also been performed, in which the n-$^{12}$Be relative energy spectra have been reconstructed by coincidence  measurements. In such a nucleus the valence neutron is weakly bound, with separation energy S$_n$=0.969 MeV, while the valence proton is strongly bound
with separation energy S$_n$=17.3 MeV. Thus the neutron will probably be emitted in the first step and then re-scattered by the core minus one proton nucleus. The projectile-target distances at which this kind of mechanism would be relevant are probably not so large to neglect the effect of the neutron-target interaction. One might wonder therefore on how to describe a neutron which breaks up because of the interaction with the  target, is  left in a plane wave moving with the same velocity of its original core and  re-interacts  with it in the final state.  This mechanism could be at the origin of the coincidence measurements  for  a one-neutron halo system like $^{11}$Be or for a projectile  like $^{14}$B.  It could be also one of the mechanisms giving rise to  $^{13}$Be in fragmentation measurements of $^{14}$Be. Supposing the two neutrons strongly correlated and being emitted simultaneously due to the interaction with the target,  the coincidence measurement of one neutron with the core would evidence the neutron-core final state interaction.

Light unbound nuclei have  attracted much attention  
\cite{bbv}-\cite{bj} in connection with exotic halo nuclei. Besides, a precise
understanding of unbound nuclei is essential to determine the position  
of the driplines in the nuclear mass chart. In two-neutron halo nuclei  
such as $^{6}$He, $^{11}$Li, $^{14}$Be,  the two neutron pair  
is bound,
although weakly, due to the neutron-neutron pairing force, while each  
single extra neutron is unbound in the field of the core.
In a three-body model these nuclei are described as a core plus two  
neutrons. The properties of core plus one neutron system are
essential and structure models rely on the knowledge of angular  
momentum and parity as well as energies and corresponding neutron-core
effective potential, therefore spectroscopic strength for neutron  
resonances in the field of the core. Ideally one would like to study
the neutron elastic scattering at very low energies on the ''core"  
nuclei. This is however not feasible at the moment as many such
cores, like $^{9}$Li, $^{12}$Be or $^{15}$B are themselves unstable and  
therefore they cannot be used as targets. Other indirect
methods instead have been used so far, mainly aiming at the  
determination of the energy and angular momentum of the continuum  
states.

Unbound nuclei have been created in several different ways besides the projectile fragmentation  \cite{nig,jl,Usa,bjorn}-\cite{6he} mentioned above: multiparticle transfer  
reactions \cite{Ale83}-\cite{Belo98} or just one proton \cite{sf,cha}  
stripping. In
a few other cases the neutron transfer from a deuteron  
\cite{Kor95}-\cite{bj} has been induced and the neutron has undergone a  
final
state interaction with the projectile of, for example $^{12}$Be. In  
this way the $^{13}$Be resonances have been populated in what can
be defined a ``transfer to the continuum reaction"  
\cite{bb}-\cite{bb4}. Thus the neutron-core interaction could be  
determined in a
way which is somehow close in spirit to the determination of the  
optical potential from the elastic scattering on normal nuclei.
In both the projectile fragmentation or the transfer method the  
neutron-core interaction that one is trying to determine appears in
the reaction as a "final state" interaction and therefore reliable  
information on its form and on the values of its parameters can
be extracted only if the primary reaction is well under control from  
the point of view of the reaction theory.

In a recent paper \cite{bbv} we showed that among the methods discussed  
above to perform spectroscopy in the continuum, the neutron
transfer method looks very promising since the reaction theory exists  
and it has been already tested in many cases \cite{bbv},
\cite{bb}-\cite{bb4}. It is important to remember that the final state  
interaction of the neutron with the target (or with the projectile, in  
the case of inverse kinematics reactions) is contained in the transfer  
to the
continuum method developed in Refs.\cite{bb}-\cite{bb4}.
 
 In this paper  and in particular in Sec. 2 the basic formalism  
to describe projectile fragmentation, an inelastic-like
excitation to the neutron-core continuum \cite{adl,bw}, is  
presented
and the effect of final state interaction of the neutron with the  
projectile core is studied. The model  is a  theory which would  
then be
relevant  to the interpretation of neutron-core coincidence  
measurements in nuclear elastic breakup reactions. In the present work we  apply it to the breakup of the  halo nuclei $^{11}$Be,  $^{14}$B but also $^{14}$Be. In the case of two nucleon breakup we try to describe here only the step in which a neutron is knocked out from the projectile by the neutron-target interaction to first order and then re-interacts in the final state with the core. The case in which a resonance is populated by a sudden process while the other neutron is stripped has been already discussed in Ref.\cite{bhe}  and we will show that there is   a simple link with the model presented here. The present model is therefore partially related to Ref.\cite{bhe}. We assume that the neutron which is not detected has been stripped while the other suffers an elastic scattering on the target. But while in Ref.\cite{bhe} the so-called transparent limit was used for the second neutron, corresponding to no interaction at all between the neutron and the target, we will consider here explicitly the effect of such an interaction on the n-core relative energy spectrum. This will result into  a core-target impact parameter dependence for the fragmentation form factor. However, in most of our calculations, we shall also use the no-recoil approximation for the core (cf. Fig.1). On the other hand  the influence of a possible second nucleon, when appropriate, is taken into account only by a modification
of the neutron-core interaction in the final state. A simple idea for relating  the present work to its future development into a two nucleon breakup model is presented in Sec. 3.  
   Section 4 contains the results of our numerical calculations
for $^{11}$Be  which, being already well understood,  has been used here  as  a test case. It also summarizes  experimental results and the present theoretical  
understanding of  $^{13}$Be.  Furthermore details on our assumptions for the potentials needed in the calculations are  presented.  Numerical results for $^{13}$Be  are contained in Sec. 5. Finally our conclusions are contained in Sec. 6.

 \section {Inelastic excitation to the continuum. }

\begin{figure}[h!b]
         \scalebox{0.3}{
              \includegraphics{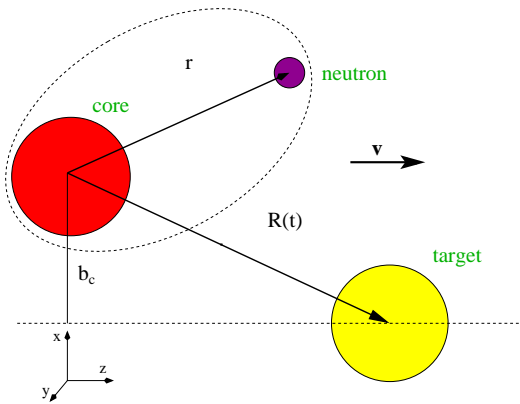}}  
              \caption{Coordinate system used in the calculations}                  
  \label{refsys}     \end{figure}
To first order the inelastic-like excitations can be described by the  
time dependent
perturbation amplitude  \cite{bb1,adl,bw}:

\begin{equation}A_{fi}={1\over i\hbar}\int_{-\infty}^{\infty}dt \langle \psi_{f} ({\bf r},t)|V_2({\bf { r-R}}(t))|\psi_{i}({\bf r},t)\rangle,\label{1}
\end{equation}
for a transition from a nucleon bound state $\psi_i$ to a final state  
$\psi_f$ which can be a bound state or a continuum state. In this paper we shall treat   
only continuum final states.
 $V_2$ is the interaction responsible for the neutron transition (cf. Eq. (2.15) of \cite{bb1}).  The potential $V_2({\bf { r-R}}(t))$
 moves past on a constant velocity path with velocity $v$ in the z-direction with an impact parameter $b_c$ in
the x-direction in the plane $y = 0$. These assumptions and the other discussed in Sec. 2.4 make our semiclassical model valid at beam energies well above the Coulomb barrier. This is in fact the regime in which projectile fragmentation experiments are usually performed (cf. Sec. 4).
The coordinate system used in the calculations is shown in Fig.\ref{refsys} and it corresponds to the no-recoil approximation for the core. In the case of the very weakly bound $^{11}$Be we will drop this approximation and explicitly take into account core recoil by defining {\bf R}(t) as the projectile-target relative motion coordinate.

Let  $\psi_{i} ({\bf r},t)=\phi_{i}({\bf r})e^{-{i\over \hbar}\varepsilon_it}$ be the single particle initial state wave function. Its  radial part $ \phi_{i}({\bf r})$ is calculated in a potential $V_{WS}( r)$ (cf. Sec. 4) which is  
fixed in space.  
  In the special case of exotic nuclei
the traditional approach to inelastic excitations needs to be modified.   
  For example the final state can be eigenstate of a potential $V_1$
modified with respect to $V_{WS}$  because some other particle is  
emitted during the reaction process as discussed in the introduction. The final state interaction might also have  
an imaginary part which would take into account the  
coupling between
a continuum state and an excited core. For these reasons our treatment is  closer to the formalism used in Ref.\cite{bw} to treat rearrangement collisions  between heavy ions than to the inelastic excitation formalism of  \cite{adl}. For initial and final states of different  angular momentum  our wave functions are trivially orthogonal due to the orthonormality of their angular parts. For transitions conserving the angular momentum of the single particle states we can orthogonalise the initial and final states as suggested  by  Eq. (8), pag. 303,  Ch.V.3, of  Ref.  \cite{bw}. Then
\beq
{\phi_{f}}^{orth}=\phi_{f}-\langle \phi_{i}\vert\phi_{f}\rangle \phi_{i}, \label{orth}\eeq
and the first order time dependent perturbation amplitude  reads
\begin{equation}
A_{fi}={1\over i\hbar }
\int_{-\infty}^{\infty}dtd{\bf r} ~ \phi^*_{f} ({\bf r})\phi_{i}({\bf r}){e^  
{i\omega t}}V_2({\bf { r-R}}(t))-\Delta A_{fi},\label{a}
\end{equation}
where $\hbar \omega$ is the energy difference between the initial and  
final states and 

\beq\Delta A_{fi}={1\over i\hbar } \langle \phi_{f}\vert\phi_{i}\rangle
\int_{-\infty}^{\infty}dtd{\bf r} ~ \phi^*_{i}( {\bf r})\phi_{i}({\bf r}){e^  
{i\omega t}}V_2({\bf { r-R}}(t)).\eeq
Because of the displacement of V$_2$ with respect to $\phi_{i}$ and in particular for the choice of a $\delta$-potential  discussed below,  it is  expected that the correction due to $\Delta A_{fi}$ will be small. Therefore we shall neglect it in the following.

Now change variables and put $z - vt = z^{\prime}$ or $t = (z -  
z^{\prime})/v$.  The excitation amplitude becomes

\begin{equation}
A_{fi}={1\over i\hbar v}
\int_{-\infty}^{\infty}dxdydzdz^{\prime} ~ \phi^*_{f}  
(x,y,z)\phi_{i}(x,y,z){e^ {iq(z -  
z^{\prime})}}V_2(x-b_c,y,z^{\prime}),\label{b}
\end{equation}
where

\begin{equation}
q={{\varepsilon_f-\varepsilon_i}\over {\hbar v}}.\label{q}
\end{equation}
Then

\begin{equation}
A_{fi}={1\over i\hbar v}
\int_{-\infty}^{\infty}dx dy dz ~ \phi^*_{f}
(x,y,z)\phi_{i}(x,y,z){e^{iqz}}\tilde V_2(x-b_c,y,q),\label{c}
\end{equation}
where

\begin{equation}
\tilde V_2(x-b_c,y,q)= \int_{-\infty}^{\infty}dz V_2(x-b_c,y,z)e^{iqz}.\label{d}
\end{equation}
In our approach the presence of the target represented by this interaction has the effect of perturbing the initial bound state wave function and allow the transition to the continuum by transferring some momentum to the neutron. For this purpose, although
the potential $V_2(r) $
has a radius of the order of the potential of the target, it is enough to choose a simplified form of the interaction.  Therefore we
choose 
 $V_2(r) $ to be a delta-function potential
$V_2(r) = v_2\delta(x)\delta(y)\delta(z)$, with $v_2\equiv$ [MeV fm$^3$]. Then the integrals over x and y can be calculated giving
\begin{equation}
A_{fi}={v_2\over i\hbar v}
\int_{-\infty}^{\infty}dz ~ \phi^*_{f} (b_c,0,z)\phi_{i}(b_c,0,z){e^  
{iqz}}.\label{1bis}
\end{equation}
The value of the strength $v_2$ used in the calculation is  discussed in Sec. 5 and in  Appendix A.
From the above equation it is clear what the effect of the n-target $\delta$-interaction is in a time dependent approach: while in the sudden approach the initial and final state overlap is taken in the whole coordinate space, irrespective of the target and of the beam velocity, here 
the overlap of the initial and final wave functions is taken at the core-target impact parameter distance on the x-direction which is along the  distance of closest approach. The y component is zero (neutron emitted on the reaction plane preferentially) and the z-component, being along the relative velocity axis is boosted by a momentum $q$. 

 The delta-function potential should be a good approximation if $\gamma {\rm R}_T<<1$ and ${\rm kR}_T<<1$ where R$_T$ is the radius of the target, $\gamma={\sqrt{-2m  
\varepsilon_i}/ \hbar}$ is the decay length of the initial single particle wave function corresponding to the nucleon binding energy 
$\varepsilon_i$ and $k={\sqrt{2m  
\varepsilon_f}/ \hbar}$ is the nucleon final momentum in the continuum state. In this case the initial and final wave functions are rather constant over the volume of the target and can be replaced by their values at the center.  The second condition is  related to the first because the cross section becomes small if $k$ is large compared to $\gamma$ (see Sec. 2.3). We also require that the reaction should be peripheral in the sense that $ {\rm R}_c+ {\rm R}_T<{\rm b}_c$ where R$_c$ is the projectile core radius. The first condition $\gamma {\rm R}_T<<1$ means that the projectile should be a good halo nucleus. Another situation where the delta-function potential is a good approximation is when $\gamma  {\rm R}_T>>1$.  That is the initial state should be strongly bound and the initial state wave function decays rapidly inside the target. Then the $\delta$-potential should be located at the surface of the target.  If, for example, V$_2$(r) is a square well potential with radius R$_T$ and if $\gamma  {\rm R}_T>>1$, then  Eq.(\ref{1bis}) can still be used to estimate the breakup with the following changes:
\begin{itemize}
\item b$_c$ is replaced by b$_c- {\rm R}_T$ i.e. the interaction is located at the surface of the target.
\item $v_2$ is replaced by $\bar v_2$ where
\begin{equation}
\bar v_2={3\over 2} {v_2\over (\gamma-ik)^2R_T^2}.\label{ap1} \end{equation}
\end{itemize}
Thus the strength is reduced and there is an extra phase. The derivation of Eq. (\ref{ap1}) is presented in  Appendix A.

First we study the simple case where the initial  bound state and the  
final continuum state   have $l_i=l_f=0$,
then

\begin{equation}
\phi_{i}(b_c,0,z)=-{C_i\over \sqrt{4\pi}}{e^ {-\gamma r}\over r},
\label{e}\end{equation}

\begin{equation}
\phi_{f}(b_c,0,z)={{C_f \over\sqrt{4\pi}}i{k\over 2}}(h^{(-)}_0(kr)-S  
h^{(+)}_0(kr)).\label{9}
\end{equation}
These expressions are  the  
asymptotic forms of the initial and final state wave functions. Their  
use can be justified when the impact parameter is sufficiently large  
\cite{bhe} and
$r=\sqrt{b_c^2+z^2}$. $\gamma$ and
$k$ are the neutron momenta in the  
initial and final states already defined. $C_i$ is the asymptotic normalization constant of the initial state wave function while  $C_f=\sqrt{2/L}$ is the
normalization constant for the  final state. L  
is a large box radius used to normalize the continuum wave function  
(cf. Eq. (2.5) of Ref. \cite{bb}).
The quantity S is the S-matrix representing the final state interaction  
of the neutron with the projectile core. Then

\begin{equation}
A_{fi}=-{v_2\over \hbar v}{C_i C_f \over 8 \pi}
\int_{-\infty}^{\infty}dz ~ {e^ {-(\gamma -ik) r}-S^*e^ {-(\gamma  
+ik)r}\over r^2}
\cos qz.\label{2}
\end{equation}
Let us define

\begin{equation}
I_R=Re \int_{-\infty}^{\infty}dz ~ {e^ {-(\gamma - ik) r}\over r^2}
\cos qz,\label{3}
\end{equation}
and

\begin{equation}
I_I=Im \int_{-\infty}^{\infty}dz ~ {e^ {-(\gamma - ik) r}\over r^2}
\cos qz,\label{4}
\end{equation}
such that:

\begin{equation}
I(k,q)=I_R+iI_I=|I|e^{i\nu}\label{4bis}
\end{equation}
while

\begin{equation}
\bar S=S e^{2i\nu}=e^{2i(\delta+\nu)}
\label{soff}\end{equation}
then

\begin{equation}
A_{fi}=C( I-S^*I^*)\label{5bis}
\end{equation}
and

\begin{equation}
|A_{fi}|^2=C^2|I|^2|1-\bar S|^2.\label{5tris}
\end{equation}
Where now $C=-{v_2\over \hbar v}{C_i C_f \over 8 \pi}$.

\subsection {Wave functions for general $l$}

In the general case of a $l_i>0$ initial state the amplitude Eq.(\ref{1})  
has to be calculated numerically using, for example, the following
forms for the wave functions. For the initial bound state

\begin{equation}
\phi_{i}({\bf r})=-C_ii^{l_i}\gamma h^{(1)}_{l_i}(i\gamma  
r)Y_{{l_i},{m_i}}(\theta,\phi).\label{f}
\end{equation}
Due of the strong core absorption discussed in Sec. 2.4 and to get a  
simple insight at the physics of unbound nuclei, we use in
this paper the asymptotic form of the initial state wave function,  
however the exact wave function, numerical solution of the bound
state Schr\"odinger equation can be used without introducing further  
complexity in the calculations.

For the final continuum state

\begin{equation}
\phi_{f}({\bf r})=C_f k{i\over 2}(h^{(-)}_{l_f}(kr)-S_{l_f}  
h^{(+)}_{l_f}(kr))Y_{{l_f},{m_f}}(\theta,\phi).\label{p2}
\end{equation}
As it was
shown in Ref.\cite{bb}, in the case of narrow isolated resonances the  
treatment of the continuum states via the S-matrix is equivalent
to the R-matrix formalism.

\subsection {Probability spectrum}

The probability to excite a final continuum state of energy  
$\varepsilon_f$ is an average over the initial state

\begin{equation}
P_{in}={1\over 2l_i+1}\Sigma_{m_i,m_f}|A_{fi}|^2\label{g}
\end{equation}
and a sum over the final states. Introducing the quantization condition

\begin{equation}
kL=n\pi,\label{h}
\end{equation}
and the density of final states, according to Ref. \cite{bb}

\begin{equation}
\rho (\varepsilon_f) d\varepsilon_f={L\over \pi}{m\over \hbar^2 k}  
d\varepsilon_f,\label{i}
\end{equation}
the probability spectrum reads

\begin{eqnarray}
{dP_{in}\over d\varepsilon_f}={2\over \pi}{v_2^2\over \hbar^2  
v^2}{C_i^2 }{m\over\hbar^2k}{1\over 2l_i+1}\Sigma_{m_i,m_f}
 |1-\bar S_{m_i,m_f}|^2 |I_{m_i,m_f}|^2, \label{8}
\end{eqnarray}
where now \begin {equation}  |I_{m_i,m_f}|^2=\left|\int_{-\infty}^{\infty} dze^{iqz}i^{l_i} \gamma h^{(1)}_{l_i}(i\gamma  
r)Y_{{l_i},{m_i}}(\theta,0)  k{i\over 2}h^{(-)}_{l_f}(kr)Y_{{l_f},{m_f}}(\theta,0)\right|^2.\label{8bis} \end{equation} 

For simplicity the equations in this section are obtained without  spin variables in the initial and final states. The generalization including spin is given in Appendix B. 
\subsection{Approximate evaluation of the integral $I(k,q)$}

In order to study the qualitative effects of the final state  
interaction we proceed now to an approximate evaluation of the integral  
$I(k,q)$ for $l_i=l_f=0$. However the calculations presented in Sec. 5 use the exact  
integrals.
For large impact parameters $b_c$, write $r=\sqrt{b_c^2+z^2}\approx  
b_c+z^2/2b_c$. Then
\begin{eqnarray}
I(k,q)&\approx& {1\over  
b_c^2}e^{-(\gamma-ik)b_c}\int_{-\infty}^{\infty}dz~e^ {-(\gamma - 
ik)z^2/2b_c} \cos(qz)\nonumber \\
       &=& {1\over b_c^2}\sqrt{2\pi  
b_c\over(\gamma-ik)}e^{-(\gamma-ik)b_c} \exp \left(-{b_cq^2 \over  
2(\gamma-ik)}\right).
       \label{14}
\end{eqnarray}
Hence the phase $\nu $ will be given by

\begin{eqnarray}
\nu&=&-{1\over 2}arg(\gamma-ik) + kb_c-{kb_cq^2\over  
2(\gamma^2+k^2)}\nonumber \\ &=&-\footnotesize {1\over 2}arg(\gamma-ik) + kb_c \left (1-{\gamma^2+k^2 \over 8{\bar k}^2}\right)\label{j}
\end{eqnarray}
where ${\bar k }=mv/\hbar$ and we have  used Eq.(\ref{q}) to obtain $q=(\gamma^2+k^2)/2 \bar k$.
The estimated value of $|I|^2$ is

\begin{eqnarray}
|I|^2&=& {1\over b_c^4}{2\pi b_c\over \sqrt{\gamma^2+k^2}}e^{-2\gamma  
b_c}\exp \left (-\gamma b_c{\gamma^2+k^2\over 4
{\bar k}^2} \right)\nonumber \\
&=& {1\over b_c^3}{2\pi \over \sqrt{\gamma^2+k^2}}\exp \left (-2 \gamma  
b_c\left (1 +{q\over 4
{\bar k}}\right)\right).\label{k}
\end{eqnarray}
The above analytical expressions are accurate to within 10$\%$ for impact parameters around the strong absorption radius and for neutron-core energies less that 1.5MeV. The agreement improves for larger impact parameters. 

The approximate formulae  give rise to simple physical  
interpretations. The first is that we have an explicit expression for  
the dependence of $\nu $ on the neutron wave number $k$, the  
core-neutron impact parameter $b_c$ and the adiabaticity parameter $q/ {\bar k}$.  
We will discuss some of the effects of the new phase $\nu$  in Sec. 5. $|I|^2\sim {e^{-2\gamma b_c}\over b_c^3}$ 
 can be interpreted as an inelastic-like form
factor and it is interesting to compare it to the transfer to the  
continuum form factor ${e^{-2\eta b_c}\over 
b_c}$ given in Ref.\cite{bb}. The inelastic form factor decreases with  
the impact parameter much faster than the transfer form
factor. This is a  well known characteristic for final bound states  
\cite{bw} and it is 
interesting to see that it persists for final continuum states.  
Furthermore the slope parameters are in both cases
given in lowest order by the initial state decay length $\gamma$.

Finally we make connection with the sudden approximation formula Eq.(20) of Ref.\cite{bhe} which describes the  second step of a
two neutron breakup reaction as a resonance decay, when the first neutron has been stripped.  
In our notation it reads

\begin{eqnarray}
{d\sigma\over {d{{\varepsilon_f}} }}&\sim & {1\over k(\gamma^2+k^2)}\left  
({{k\cos\delta+\gamma \sin \delta}\over \sqrt {\gamma^2+k^2} }\right
)^2\nonumber \\
&\sim & {1\over k}{|\sin(\delta +\beta)|^2\over (\gamma^2+k^2)}\label{geo}
\end{eqnarray}
where  $\beta=arctan (k/\gamma)$. This formalism also predicts the  
presence of an extra phase shift $\beta$ with respect to the free
particle scattering determined by $\delta$ since $|\sin(\delta +\beta)|^2={1\over 4}|1-\tilde S|^2$ and $\tilde S=e^{2i(\delta+\beta)}$. Similarly to our case the  
effect of $\beta$ would be to modify the resonance-like structures. In  
both cases then $\bar S$  and $\tilde S$ could be interpreted as   
off-the-energy-shell S-matrices. On the other hand our additional phase $\nu$
 contains an explicit dependence on the impact parameter  and 
we calculate the potential phase shift  and S-matrix by an optical model  code.
Our S-matrix can in principle be complex to allow for  
core excitation effects. Also it can consistently, and in the same formalism,  describe resonant and non-resonant final
continuum states of  angular momentum $l_f=0$ but also $l_f>0$. In the latter case, the Breit-Wigner assumption for the line shape of the resonances used in   other approaches, is naturally given  by the optical model calculation of the factor 
$ |1-\bar S|^2$ in  Eq.(\ref{8}).

\subsection{Cross section }

In \cite{bw} it was shown that the semiclassical treatment of  
peripheral quasi-elastic reactions is valid for transfer reactions as
well as for (inelastic) projectile excitation and therefore we will  
apply it in the following by simply substituting $ P_{in}(b_c)$
to $P_{t}(b_c)$ in the well known formula which gives the cross section  
in terms of the neutron excitation probability and the core
elastic scattering probability. A full description of the treatment of  
the scattering equation for a nucleus which decays by single
neutron breakup following its interaction with another nucleus, can  
also be found in Refs. \cite{bb1,bb,bb4,jer} where the cross section
differential in $\varepsilon_f$, the final, continuum, neutron energy  
is given as

\begin{equation}
{d\sigma_{-1n}\over {d{{\varepsilon_f}} }}=C^2S
\int d{\bf b_c} {d P_{in}(b_c)\over
d{{\varepsilon_f}}}
P_{ct}(b_c), \label{cross}
\end{equation}
(see Eq. (2.3) of \cite{bb4}) and C$^2$S is the spectroscopic factor  
for the initial state.

The core survival probability $P_{ct}(b_c)=|S_{ct}|^2$ \cite{bb4} in  
Eq.(\ref{cross}) takes into account the peripheral nature of
the reaction and naturally excludes the possibility of large overlaps  
between projectile and target. $P_{ct}$ is defined in terms
of a S-matrix function of the core-target distance of closest approach  
$b_c$. 
A simple parameterisation is
$P_{ct}(b_c)=e^{(-\ln 2 exp[(R_s-b_c)/a])}$ \cite{bb4}, where the  
strong absorption radius R$_s\approx1.4(A_p^{1/3}+A_t^{1/3})$ fm is  
defined
as the distance of closest approach for a trajectory that is 50\%  
absorbed from the elastic channel and a=0.6 fm  is a diffuseness
parameter. The values of R$_s$ thus obtained agree within a few percent  
with those of the Kox parameterization \cite{kox}.

Because  $P_{in}(b_c)$ depends on $b_c$ trough  the form factor 
$I_{m_i,m_f}$, the final cross sections will get the main contribution from a limited range of
 impact parameters around the strong absorption radius.  As we shall see in the following (cf. figs.7 and 8) this localization   makes the tails of the spectra from Eqs.(\ref{8}) and (\ref{cross}) to decay faster than Eq.(\ref{geo}) towards neutron-core high energies. 

\section{Two-nucleon breakup}
 This section introduces the basis for  a generalization of Eq.(\ref{cross}) which will be further expanded elsewhere. The goal is to take into account explicitly the presence of a second, stripped, nucleon as in the cases of projectiles like $^{14}$Be which have a two-neutron halo or $^{14}$B in which a proton can also emitted.
 
\begin{figure}[h!t]
         \scalebox{0.3}{
              \includegraphics{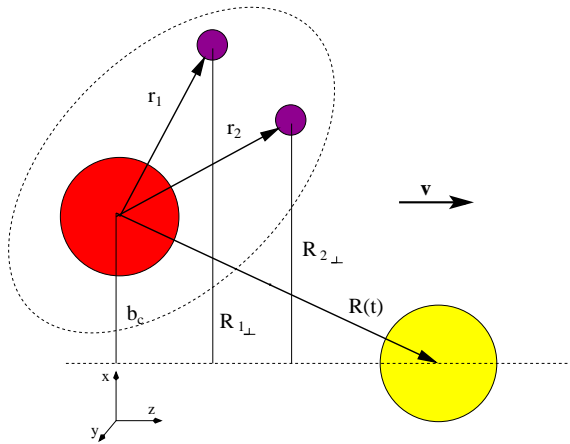}}  
              \caption{Coordinate system used in the calculations of two-nucleon breakup.}                  
  \label{duen}     \end{figure}

We start from hypothesis similar to those leading to Eq.(7) of Ref.\cite{bhe} which gives, in the case of two-neutron breakup, the one-neutron-core relative momentum distribution when the other neutron is stripped. Let us call (1) the stripped neutron and (2) the neutron detected in coincidence with the core. Following the approximations  proposed by Ref.\cite{bhe} for the coordinate variables, shown in Fig.\ref{duen},
of the core and neutrons with respect to the target one gets:  R$_{1\perp}$=r$_{1\perp}$+b$_c$ and R$_{2 \perp}$=r$_{2\perp}$+b$_c$, where   the heavy-core approximation has been used. r$_1$ and r$_2$ are  the coordinates of neutron (1) and (2)  with respect to the core, while R$_1$ and R$_2$ are with respect to the target.
  
   Suppose  the initial two-neutron wave function  to be given by a product
of  single particle wave-functions, as in the shell model: $\Psi ({\bf r}_1,{\bf  r}_2)= a_1 [\phi_1 ({\bf r}_1)
 \phi_2 ({\bf r}_2)] $
with spectroscopic factor $a_1^2=C^2S$. For simplicity we consider here only a $l_i=l_f=0$ transition and thus we do not include  spin wave functions. Then  the neutron-core cross section differential in the relative energy is:
\beq
\frac {d \sigma_{-2n}}{d\varepsilon_f }=2C^2S
\int d^2 {\bf b}_c |S_{ct}(b_c)|^2 \frac {d P_{in}(b_c)}
{d \varepsilon_f} \int d^2{\bf r}_{1\perp}
(1-|S_{1}({\bf R}_{1\perp})|^2 )\int dz  |\phi_1 ({\bf r}_{1\perp},z)|^2 . \label{cross12}
\eeq
Where  $d P_{in}(b_c)/{d \varepsilon_f}$ is given by Eq.(\ref{8})
and, similarly to  Eq.(\ref{cross}), we  are treating the core-target interaction in the eikonal approximation. The $d^2{\bf r}_{1\perp}$ integral above gives the neutron (1) stripping probability. For the S$_1$-matrix we consider a sharp cut-off approximation such that $S_{1}({\bf R}_{1\perp})=0$ if $0<  {\rm R}_{1\perp}<  {\rm R}_T$, while $S_{1}( {\rm R}_{1\perp})=1$ if R$_{1\perp}>  {\rm R}_T$ and R$_T$ is the target radius.

Thus we obtain a simple expression for the two nucleon breakup cross section, in which one is stripped by the target while the other is elastically scattered and interacts with the core in the final state 

\beq
\frac {d \sigma_{-2n}}{d\varepsilon_f }=2 C^2S
\int d^2 {\bf b_c}|S_{ct}(b_c)|^2 \frac {d P_{in}(b_c)}
{d \varepsilon_f}   \int d^2{ \bf r}_{1\perp}
\int dz  |\phi_1 ({\bf r}_{1\perp},z)|^2, \label{cross123}
\eeq
 as a product   of the neutron (2)-core relative energy distribution and a factor depending on the stripped neutron (1) wave function.  For each core-target  impact parameter b$_c$ the limits of the  integral on  r$_{1\perp}$  are: b$_c- {\rm R}_T<r_{1\perp}<{\rm b}_c$.


\section{Applications }
\subsection{The reaction $^{11}$Be $\to$ n+$^{10}$Be}

As a test of our model we calculate the relative energy spectrum  n+$^{10}$Be obtained by the authors of Ref.\cite{fuku}
in the breakup reaction of $^{11}$Be on $^{12}$C at 70 A.MeV. The structure of $^{11}$Be is  well known: the valence neutron is bound by 0.503 MeV; the wave function is mainly a 2s state with a spectroscopic factor around 0.8 and there is also a small d$_{5/2}$ component. The main d$_{5/2}$ strength is in the continuum centered around 1.25 MeV \cite{simone}. We have  calculated the initial wave function for the s-state in a simple Woods-Saxon potential with strength fitted to the experimental separation energy and whose parameters are: r$_0$=1.25 fm, a=0.8 fm. As possible final states we have considered only the s, p and d partial waves calculated in the $l$-dependent potentials of Table \ref{tb1}. The delta-function potential strength
has been chosen as -4057.59 MeV fm$^3$. The authors of Ref.\cite{fuku} have shown that the effect of Coulomb breakup is noticeable in  their  n+$^{10}$Be spectrum. We have also included this contribution, calculating it with the method of Margueron et al. \cite{jer} which explicitly takes into account the recoil of  the core. 
\begin{figure}[h!t]
\center
 \vskip 20pt
 \scalebox{0.4}{
               \includegraphics{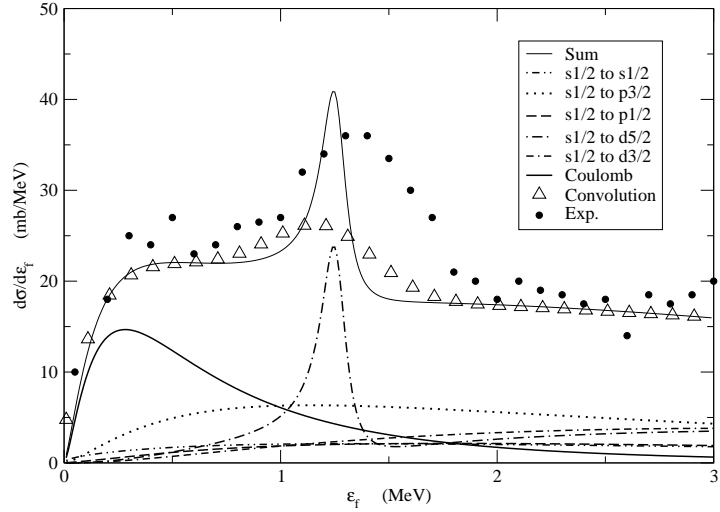}}
\caption{ \footnotesize n-$^{10}$Be relative energy  spectrum,
including Coulomb and nuclear breakup for the reaction  $^{11}$Be+$^{12}$C $\to$ n+$^{10}$Be+X at 69 A.MeV. Only the contributions from an s initial state with  
spectroscopic factor C$^2$S= 0.84 are  calculated. The  triangles are the total calculated result after convolution with the experimental resolution function. The dots are the experimental points from \cite{fuku}. }
\label{11Be}
\end{figure}

\begin{table}[h]
 \caption {\footnotesize Woods-Saxon potential parameters for the s, p and d states in the $^{11}$Be continuum.}
 \begin{center}
\begin{tabular}{c|ccc|ccc}
\hline\hline\
&$l$=0&$l$=1&$l$=2 & V$_{so}$(MeV)&a (fm)& R (fm)  \\
\hline
V$_0$(MeV) & -62.52     & -39.74      &   -63.57  &5.25    & 0.6     &2.585        \\
\hline\hline
\end{tabular}
\end{center}
\label{tb1}
\end{table}

The spectrum of Fig.\ref{11Be} is very similar to the spectrum obtained in Ref.\cite{cap} by solving 
the time-dependent Schr\"odinger equation  numerically,  expanding 
the projectile wave function upon a three-dimensional spherical mesh. Similarly to the present model, a classical, straight line trajectory for the core-target scattering was used in Ref.\cite{cap}. Also our n-core potentials are very close to those
of Ref.\cite{cap} and our $\delta$-interaction strength is consistent with the volume integral of their neutron-target interaction.

We have then  folded the calculated spectrum througth the experimental resolution function of  Fukuda et al. \cite{fuku}, as given in Ref.\cite{cap} 
\begin{eqnarray}
{d\sigma_{-1n}^{conv}\over {d\varepsilon}}=\int {d\sigma_{-1n}^{theo}\over {d \varepsilon_f}}g(\varepsilon_f-\varepsilon  ) d\varepsilon_f \nonumber\\
g(\varepsilon_f-\varepsilon)={{1}\over{0.48\sqrt{\varepsilon_f}}}
\exp{\left(-{(\varepsilon-\varepsilon_f )^2\over 0.073 \varepsilon_f} \right)}
\end{eqnarray}
with $\varepsilon_f$ the energy in the theoretical calculation. The result is shown in Fig.\ref{11Be} by the triangles. The full curve is the total spectrum, sum of Coulomb and nuclear breakup. Each individual transition, due to the nuclear interaction only, is also shown.
The dots are the experimental points from \cite{fuku}. Our calculations include an initial state spectroscopic factor C$^2$S=0.84. The kind of discrepancy between  our calculation and the data in the range 1-2 MeV is very similar to that of the calculations in Ref.\cite{cap}.

Encouraged by the good agreement of  our results  with those of Refs.\cite{fuku} and \cite{cap}, we conclude that our model is quite reliable to calculate projectile fragmentation including final state interaction with the core and we turn now to the study of  more challenging reactions.


\subsection {The reactions $^{14}$Be$\to$$^{13}$Be +n and  
$^{14}$B$\to$$^{13}$Be +p }

An interesting unbound nucleus which has recently attracted much attention is $^{13}$Be.
It   has been obtained in several different type of experiments with normal and exotic beams but its structure is not  clear yet.

One of the aims of this paper is to see whether the neutron-$^{12}$Be  
relative energy spectra obtained
from fragmentation of $^{14}$Be or $^{14}$B would show differences  
predictable in a theoretical model. Therefore we start by describing  
briefly the present knowledge of  $^{13}$Be.

\subsection{ Structure of $^{13}$Be }

   The first experimental evidence of $^{13}$Be was recorded in Ref.\cite{Ale83}. It was the unobserved particle in the two-body reaction $^{14}$C($^{7}$Li,$^{8}$B) at 82 MeV. A narrow resonance at 2 MeV above the neutron emission threshold was observed in the spectrum
of the measured $^8$B ions and it was interpreted as being due to the ground state of $^{13}$Be. Later on it was identified as a d$_{5/2}$ state  in the double-charge exchange experiment
$^{13}$C($^{14}$C,$^{14}$O) at E$_{inc}$=337 MeV \cite{Ostr92} and in an inverse kinematics (d,p)  transfer reaction at 55 A.MeV \cite{Kor95}: both the missing mass method, from the 
detected proton spectrum, and the invariant mass spectroscopy, from the measurement of all the decay products from the unbound state were used.
  The d$_{5/2}$ resonance was considered as the ground state of $^{13}$Be until the
experiment of Ref. \cite{Belo98}. This experiment used a stable beam multi-nucleon 
transfer process  $^{14}$C($^{11}$B,$^{12}$N)$^{13}$Be in which a lower state, unbound by 800
keV was observed. A spin   J=1/2 was suggested but without parity assignment.  More recently
 a broad peak has been obtained in several projectile fragmentation experiments
\cite{jl,Thoen00,simo,nak} and tentatively  identified as a  1/2$^+$ state.  The experiment  of Ref.\cite{Thoen00} used fragmentation of $^{18}$O at 80 A.MeV. Neutrons in coincidence with 
$^{12}$Be were detected and the broad peak was observed in their relative velocity  spectrum.  The spectrum was fitted with a virtual s-state of energy 60 keV while in Ref.\cite{jl}, which used fragmentation of   $^{14}$B a
virtual s-state could not fit the experimental  neutron spectrum. The
assumption of an s-resonance  of energy 700keV and width  1.3 MeV
leads  to a good agreement. In Refs.\cite{jl,Usa,simo,nak} these unbound states of $^{13}$Be have been
populated by breakup of $^{14}$Be or $^{14}$B. Both types of experiments show a low
energy  peak but only in the breakup of $^{14}$B \cite{jl,Usa}   the other peak
at 2 MeV corresponding to the d$_{5/2}$ state, is seen clearly. Finally Simon et al. \cite{simo} have fitted their spectrum from $^{14}$Be breakup with s, p and d components.

On the other hand the $^{13}$Be structure is the crucial input of any three-body
model describing $^{14}$Be as two neutrons outside an inert $^{12}$Be core
\cite{Bert91,Thomp96,Vinh96}. One then needs to know the n-$^{12}$Be interaction. 
Assuming
in $^{12}$Be a normal order of shells and an unbound s-ground state  of
$^{13}$Be, the d$_{5/2}$ resonance, experimentally at 2 MeV, has to be lowered in
order to get the experimental two-neutron separation energy in $^{14}$Be
\cite{Thomp96,Vinh96}. An inversion of 
2s-1p$_{1/2}$ shells, similar to the inversion in $^{11}$Be and $^{10}$Li, was shown to solve this discrepancy \cite{jl,pa}.  This inversion is   due in part  to the coupling of the neutron with
core collective excited states,  inducing in the wave function a
 component with one neutron coupled to  the 2$^+$ state of the core. 
 Descouvemont, made a GCM calculation \cite{de,deb}. He  expands the $^{13}$Be
and $^{14}$Be wave functions as a superposition of one and two neutrons    plus
a core of $^{12}$Be, thus having contributions  of components on excited
states of the core.  He  gets an agreement with experimental values of the
binding energy of $^{14}$Be and of the d$_{5/2}$ resonance energy. His model gives  a
s-ground state in $^{13}$Be  very weakly bound (100 keV).  However due to the
uncertainties inherent to any model, this result may  not be inconsistent
with the experimental evidence for a weakly unbound $^{13}$Be.  These results were
confirmed later on by Baye and collaborators in a Lagrange-mesh calculation
\cite{ba,bab}. Recently Tarutina et al. \cite{ta}   studied $^{13}$Be and
$^{14}$Be as one and two neutrons outside a deformed core using a hyperspherical 
harmonics expansion method. They found that both $^{14}$Be and $^{13}$Be (with an
unbound   s-ground state) are well described with a large and positive
quadrupole deformation. Therefore to disentangle the various theoretical
descriptions of  $^{13}$Be and $^{14}$Be one needs more precise experimental 
information on the structure of both and in particular on their  spectroscopic
factors.

\subsection{ Structure of  $^{14}$Be and $^{14}$B }

These uncertainties in the interpretation of experimental results as  
compared to structure calculations were at the origin of our  
motivations to
try to understand  whether the neutron-$^{12}$Be relative energy  
spectra obtained
from fragmentation of $^{14}$Be or $^{14}$B  would show differences  
predictable in a theoretical model.
It is likely that if differences will be found  in the experimental results
with $^{14}$B and $^{14}$Be beams they could be  due to an interplay between structure and reaction effects. 

The ground state of $^{14}$Be  
has spin $J^\pi=0^+$. In a simple model assuming two
neutrons added to a $^{12}$Be core in its ground state the wave  
function is:

\begin{equation}
|^{14}Be>=[b_1(2s_{1/2})^2+b_2(1p_{1/2})^2+b_3(1d_{5/ 
2})^2]\otimes|^{12}Be,0^+>\label{l}
\end{equation}
Then the bound neutron can be in a 2s, 1p$_{1/2}$ or 1d$_{5/2}$ state.  
However, as it has been discussed in the previous section, the situation is much more complicated \cite{Bert91}-\cite{bab} and  in particular the
calculations of Ref. \cite{ta} show that there is a large component $(2s_{1/2},1d_{5/ 
2})\otimes|^{12}Be,2^+>$
 with the core in  
its low energy 2$^+$
state which can modify the neutron distribution. 

The ground state of $^{14}$B has spin $J^\pi=2^-$. 
In a model where it is described as a neutron-proton pair
added to a $^{12}$Be core in its 0$^+$ground state with the proton in  
the 1p$_{3/2}$
shell, its wave function may be written as:

\begin{equation}
|^{14}B>=[a_1(p_{3/2},2s_{1/2})+a_2(p_{3/2},  
d_{5/2})]\otimes|^{12}Be,0^+>\label{m}
\end{equation}

The present experimental information \cite{msu} on  $^{14}$B is that the
neutron is in a state combination of s and d-components with weights  
66\% and 30\% respectively, while shell model calculations show a similar
mixture and no component with an excited state of the core.
There are two possibilities for the reaction mechanism. One is that
a proton is knocked out in the reaction with the target. The remaining  
$^{13}$Be would be left
in an unbound  s-state with probability $| a_1 |^2$, in a  
d$_{5/2}$-state with probability
$| a_2 |^2 $.  These unbound
states would decay showing the s-wave threshold and d-wave resonance  
effects. As mentioned in the introduction, the second possibility is that the neutron is knocked out first due to its small separation energy and that the proton is stripped from the remaining  $^{13}$B. We show in Sec. 5 that this can also lead to resonance-like effects in the cross section. 

\subsection{One neutron average potential }

The link between reaction theory and structure model is made by the  
neutron-core potential determining the S-matrix in Eq.(\ref{9}).
Then if the theory fits the position and shape of the continuum  
n-nucleus energy distribution, obtained for example by a coincidence
measurement between the neutron and the core,  the parameters of a  
model potential  can be deduced.  
Our initial  bound states are obtained in a simple Woods-Saxon with R= r$_0A^{1/3}$
  \begin{equation}V_{WS}(r)={V_0\over{1+e^{(r-R)/ a} }}-\left  
({\hbar\over m_{\pi}c}\right )^2{V_{so}\over ar}{e^{(r-R)/
a}\over {(1+e^{(r-R)/ a})^2}}{\mathbf {l \cdot \sigma}}\label{n}
\end{equation}
The depth is adjusted to fit the  
binding energies given in Table \ref{o} and  Fig.\ref{fig6a}. Other parameters are also given in Table \ref{o}.

\begin{table}[h]
\caption {\footnotesize Asymptotic normalization constants C$_i$(fm$^{- {1/2}}$) for the initial state wave functions of the bound neutron.  Potential parameters are:  
V$_0$  fitted to give the  two energies -0.97 MeV  and -1.85 MeV, which  are the known neutron  
binding energies in  $^{14}$B and  in  $^{14}$Be respectively \cite{nucbase}. The  
other potential parameters are r$_0$=1.27fm,
a=0.75fm,  V$_{so}$=5.25MeV.   }
\begin{center}\footnotesize \begin{tabular}{c|c}
\hline
\hline
$\varepsilon_i $(MeV)&-0.97~~~~-1.85  
\\ \hline 

$l_i,~~ j_i$&  C$_i$(fm$^{- {1/2}}$)\\ \hline

 0 $~~{1/2}$   &1.31~~~~1.99 \\
 1 $~~{1/ 2}$   &0.55~~~~0.88 \\
2 $~~{5/2}$   &0.17~~~~0.34  \\
\hline\hline
\end{tabular}\end{center}\label{o}
\end{table}

To describe the valence neutron in the $^{13}$Be continuum we
assume that the single neutron hamiltonian with respect to $^{12}$Be  
has the form
\begin{equation}
h=t+ U+iW
\label{op}\end{equation}
where $t$ is the kinetic energy and
\begin{equation}
U (r) =V_{WS}+\delta V \label{pot}
\end{equation}
is the real part of the neutron-core interaction. For the time being the
imaginary part is taken equal to zero. V$_{WS}$ is again a Woods-Saxon  
potential plus spin-orbit whose parameters are given in Table \ref{p},
and $\delta$V is a correction \cite{Vinh96}:
\begin{equation}
\delta V(r)=16\alpha {e^{{{2(r-R)}/ a}}/({1+e^{{{(r-R)}/ a}}})^4  
}\label{pot1}
\end{equation}
which originates from particle-vibration  
couplings. They are important for low energy states but can be
neglected at higher energies.  The above form  is suggested by a calculation of such couplings using
 Bohr and Mottelson collective model of the transition amplitudes between zero and one phonon
states. Therefore our structure model is not a simple {\it single-particle in a potential} model but contains in it the full complexity of single-particle vs. collective couplings. However the fact that such a complexity can be put in a form like Eq.(\ref{pot1}) is  an added value to our approach. If simple fittings of experimental data will be obtained, then  the parameters of a semi-phenomenological potential  can be deduced and  linked to a more microscopic model. A more realistic treatment would require the description of both bound and unbound states in a three-body model such as in Refs.\cite{rod} and \cite{IJT}. 

\begin{table}[hb]
\caption {\footnotesize Woods-Saxon and spin-orbit potential
parameters for the continuum final states.}
\begin{center}
\footnotesize
\begin{tabular}{ccccc}
\hline
\hline
  V$_0$  &  r$_0$  &  a   & V$_{so}$ & a$_{so}$ \\
(MeV) &  (fm) & (fm) & (MeV)  &  (fm) \\
\hline
-39.8   &   1.27  & 0.75   &   6.9   & 0.75 \\
\hline
\hline
\end{tabular}
\end{center}\label{p}
\end{table}

Table \ref{q1}  gives  the energies and widths of the 1p$_{{\footnotesize {1/2}}}$ and  
1d$_{5/2}$ states, chosen according to
Ref.\cite{Vinh96} with
different values of the strength $\alpha$. The widths are obtained from  
the phase shift  variation near  resonance energy, according to
$d\delta_{j}/d\varepsilon_f|_{\varepsilon_{res}}=2/\Gamma_j$, once that  
the
resonance  energy is fixed \cite{joa}. We stress here that while the position of our d$_{5/2}$ resonance  agrees with the experimental evidences discussed in Sec. 4.3, the position of our 1p$_ {1/2}$ resonance is only an hypothesis \cite{Vinh96,Labi99}.
\begin{table}[ht]
\caption{\footnotesize Energies and widths of unbound p- and d-states  
in $^{13}$Be  and corresponding strength parameters for the $\delta$V  
potential.}\vskip .1in
\begin{center}
\footnotesize
\begin{tabular}{lcccc}
\hline
\hline\
&&$\varepsilon_{res}$ &$\Gamma_j$& $\alpha$\\
&&(MeV) &(MeV) & (MeV)\\
\hline
&1p$_ {1/2}$ & 0.67 &0.28 &8.34\\
&1d$_{5/2}$ & 2.0 &0.40&-2.36\\
\hline
\hline
\end{tabular}
\label{q1}\end{center}
\end{table}

\section {Results }

\begin{figure}[h]
\center
 \vskip 20pt
 \scalebox{0.4}
{\includegraphics[angle=-90]{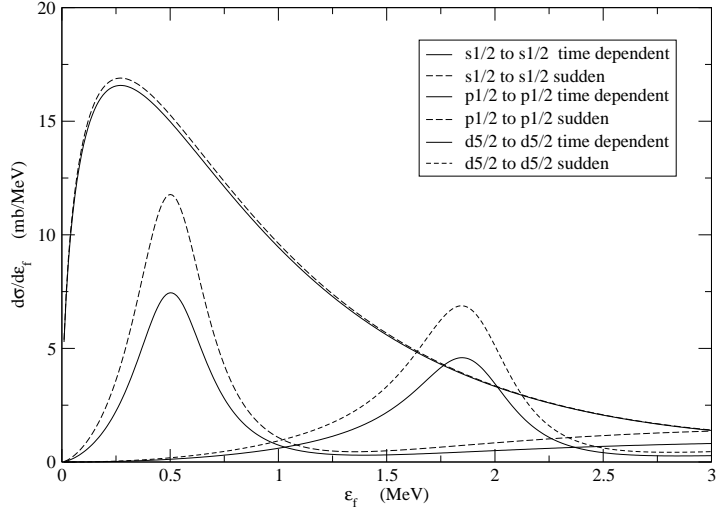}}
\caption{  Population of resonances in the n-$^{12}$Be relative energy  spectrum.  
Comparison of sudden (dashed line) and non-sudden (solid line) results  
for
an s $\to$ s transition with peak at 0.25 MeV, p$\to$ p with peak at 0.5 MeV  and d$\to$ d transition with peak around 2 MeV. See text for details.}
\label{fig2}
\end{figure}

\begin{figure}[h]
 \vskip 20pt
\center  \scalebox{0.4}
{\includegraphics[ angle=-90]{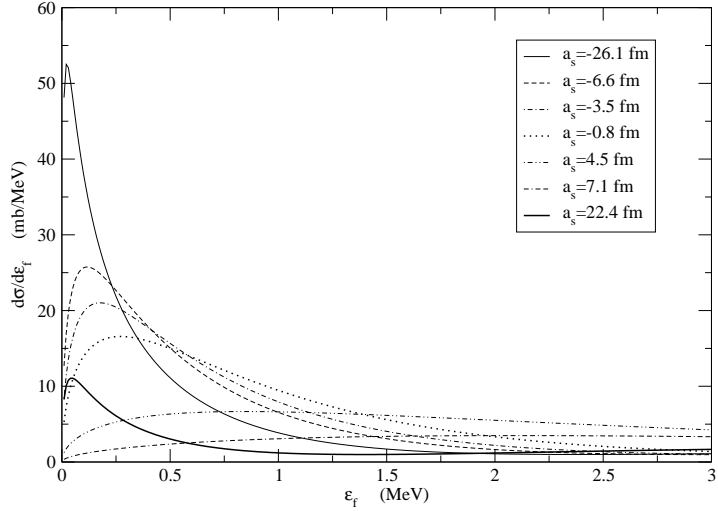}}
\caption {\footnotesize   Comparison of results obtained  
considering a final  s-state for the n-$^{12}$Be  
relative energy
spectrum with positive and negative scattering lengths. Scattering lengths are given on the figure and their  
corresponding $\delta$V potential strengths in Table \ref{r}.}
\label{fig3a}
\end{figure}

\begin{figure}[h]
 \vskip 20pt
\center  \scalebox{0.4}
{\includegraphics[ angle=-90]{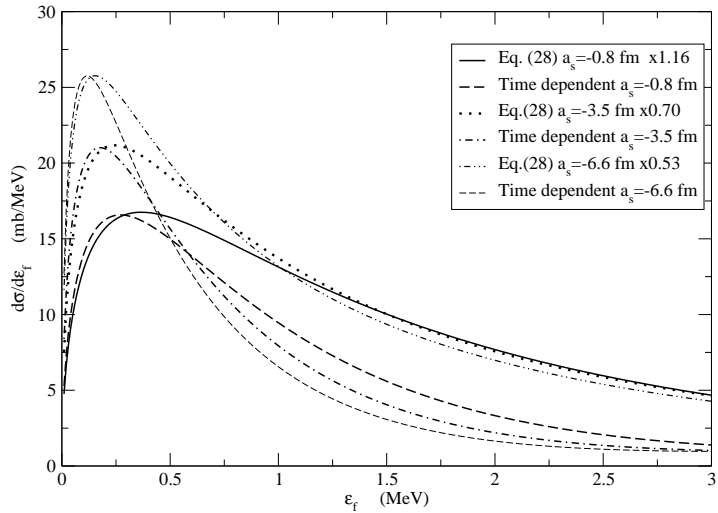}}
\caption {\footnotesize Comparison of time dependent calculation for an s to s transition with results of Eq.(28) using the same optical model phase shifts corresponding to scattering lengths as indicated. For each case we give in the legenda the normalization factor between the two calculations. }
\label{fig3b}
\end{figure}

\begin{figure}[h!t]
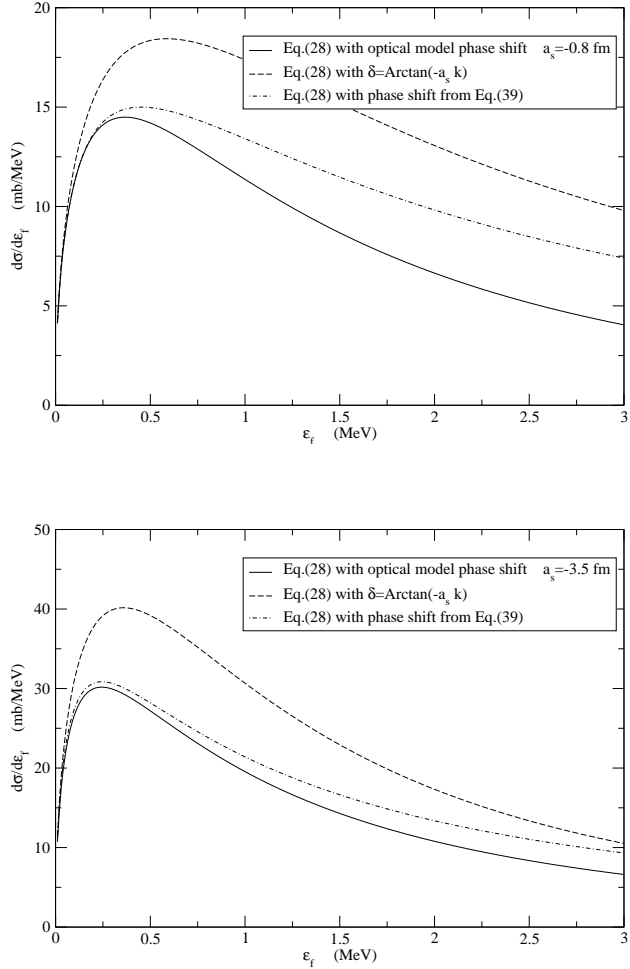
 \center  \scalebox{0.35}
{\includegraphics{Bertsch_comp_eff_mdot8.eps}}
\vskip .98cm \scalebox{0.35}
{\includegraphics{Bertsch_comp_eff_m3dot5.eps}}
\caption {\footnotesize  Results of Eq.(28) using three   different prescriptions for the phase shifts corresponding to scattering lengths as indicated. }
\label{fig4}
\end{figure}

Results obtained with the model outlined in Secs. 2  and 4  will now be  
discussed. We describe the reaction corresponding to a neutron  
initially bound in $^{14}$Be or   $^{14}$B which is then excited into  
an unbound state of $^{13}$Be, assuming that another nucleon has been  
emitted and stripped by the target, thus not detected in coincidence with the core. The sudden approximation   studied in Ref.\cite{bhe},  similar to our q=0 case,
 was found to be excellent for  energy distributions like  those discussed in  
this work. 

One of the goals of the present calculations, as far as the reaction model is concerned, is to
understand the  incident energy dependence of the breakup cross-section and the 
dependence on
the neutron initial binding energy. Related to this is  the investigation of the validity of the  
sudden approximation and the accuracy necessary in calculating the phase shifts. Finally we wish   to understand how
 the presence of p- and d-wave resonances, besides a threshold s-state in the final state, can affect the results.

As a preliminary to  a future, more accurate, study of the breakup of $^{14}$Be and $^{14}$B in a fully time dependent method, we consider the knockout of a single neutron from a bound state in a potential, similarly to the previous calculation for $^{11}$Be.
The calculations in the present section are made with different  
potentials for the initial
and final state.  The initial state is bound with a separation energy in  
the range 1-2 MeV and
  the final state is unbound. The continuum energies can be adjusted by varying the parameter  
$\alpha$ in the potential.
By changing the strength $\alpha$  
of the $\delta$V potential in Eq.(\ref{pot1}) we will make also  the  2s-state just bound  
near threshold and see
 what would be  
  the effect on the continuum spectrum. 
  The delta-interaction strength used in Eq.(\ref{d}), is $v_2$=-8625 MeV fm$^3$. It has been obtained by imposing that this interaction gives the same volume integral as a n-$^{12}$C Woods-Saxon potential of strength -50.5MeV, radius 2.9 fm and diffuseness 0.75 fm. With a diffuseness of 0.5 fm one would obtain $v_2$=-6717  MeV fm$^3$. The value $v_2$=-7481 MeV fm$^3$ would be obtained from a Woods-Saxon with the same geometry and a depth fitted to give the experimental  neutron separation energy  in $^{13}$C.

  First we study the differences between results from Eqs.(\ref{8}) and (\ref{cross}), the sudden approximation limit  q=0 of those equations and Eq.(\ref{geo}). 
In Fig.\ref{fig2}  we show the calculation of the  
differential probability for the transition from a bound s-state to an
unbound s-state, a bound p-state to an
unbound p-state and a bound d-state to an
unbound d-state. The initial state  
parameters are given in Table \ref{o}. The full lines  correspond to the case  
$\varepsilon_i$=-1.847 MeV, E$_{inc}$=70 A.MeV corresponding to
v=11.35 (fm$\times$10$^{22}$ sec$^{-1}$) and q in the range (0.025 $\to$  0.065) fm$^{-1}$. The dashed
lines give the q=0 calculations of  Eqs.(\ref{8}) and (\ref{cross}). There is a
small
difference in the absolute value of the probability and the sudden  
calculation results have slightly different widths in the s to s case. In the other two cases the differences are noticeable. We have considered only the three transitions keeping $j_i=j_f$
as it would happen in an extreme sudden transition.   
Then we
studied the effect of the extra phase in Eqs.(\ref{8}) and (\ref{geo}) on the  
position of the resonance peaks, and we found that
the shift is negligible and would not be noticeable for the range of  
neutron-core energies discussed here. Also we
have calculated for several velocities ranging from 10 to 23 (fm$\times$10$^{22}$ sec$^{-1}$) and found no noticeable
differences. On the other hand changing the initial binding energy from  
-0.97 MeV to -1.85 MeV gives a widening of
the distribution and a slight shift of the peak value, as shown in Fig.  
\ref{fig6a}. These two energies are the known neutron binding energies in   
$^{14}$B and  in  $^{14}$Be respectively \cite{nucbase}.  In this work 
 we have kept the initial separation energy fixed at the value  
1.85 MeV, unless otherwise stated.
Our conclusion is that for fragmentation reactions such as those  discussed here, the sudden
approximation q=0 in    Eqs.(\ref{8}) and (\ref{cross}), is well justified from the point of view of the  
independence from the beam velocity. On the other hand the value of the 
$\varepsilon_f-\varepsilon_i$ difference has an important effect on the results when the final energy increases and for states with $l_f>0$.

The first peak shown by the experimental spectra of $^{13}$Be needs to be studied in great detail as it would help determining the ground state angular momentum and parity. In particular, if it is due to an s-state its characteristics will depend on its closeness to threshold. Therefore we study now such a point.

\subsection{  Low energy s-states}

   Using
the effective range formula \cite{joa}

\begin{equation}
k \cot \delta_0=-{1\over a_s}+{1\over 2}r_ek^2,\label{er}
\end{equation} in the case of a bound state of small binding energy  
$\gamma \to 0$  one has

\begin{equation}-{1\over a_s}=-|\gamma|+{1\over 2}r_e\gamma^2.\label{nu}
\end{equation}
Equation (\ref{nu}) has been used  \cite{blatt} also to define an {\it  
energy} of unbound s-states of near zero energy. Such a procedure is  
reliable when $\gamma R$ is smaller than about 0.5, where $R$ is the
radius of the potential. Therefore the effective range formula is  
accurate only for very
low energies.  Thus we have fitted the behavior of our  s-state
phase shifts from the optical model calculation, to Eqs.(\ref{er}) and (\ref{nu}) and the corresponding  
parameters are given in Table  \ref{r}. In the case of a bound state, the  
effective range values can also consistently be obtained from Eq.  
(\ref{nu}).

\begin{table}[h]
\footnotesize
\caption {\footnotesize Strengths of the s-state $\delta$V potential in  
Eq.(\ref{pot1}) and corresponding scattering lengths, effective
range parameter and {\it energy parameter $\epsilon $}. The strength of the central  
Woods-Saxon part is V$_0$=-39.8 MeV in all cases (cf. Table \ref{p}). }\vskip  
.1in
\begin{center}
\begin{tabular}{lcccc}
\hline
\hline\
&$\alpha$& a$_s$ & r$_e$& $|\epsilon |$\\
&  (MeV) & (fm)  & (fm) &     (MeV)    \\ \hline
&   8.0  & -0.8  & 117.0 &              \\
&   4.0  & -3.5  & 17.9 &              \\
&   2.0  & -6.6  & 11.8 &              \\
&  -1.0  & -26.1 & 7.58 &              \\
& -5.0   &  22.4 & 5.9  &      0.06        \\
& -15.0  &  7.1  & 3.8  &      1.34        \\
& -35.0  &  4.5  & 2.7 &       6.49       \\
\hline
\hline
\end{tabular}
\label{r}\end{center}
\end{table}

Figure \ref{fig3a}  shows the influence of the phases $\delta $ and $\nu$ (cf. Eq.(\ref{soff}))  
on the breakup cross sections. 
The results shown
correspond to  final s-states with positive and negative scattering  
lengths. Several cases are considered and the corresponding potentials,  
scattering lengths, and effective
ranges are given in Table \ref{r}. The scattering length values 
were obtained from the phase shifts as $a_s=-  
\mathrel{\mathop{lim}\limits_{k\to 0\hspace{.28em}}}{tan\delta_0\over  
k} $, and also cross-checked by solving the Schr\"odinger equation at  
zero energy. The bound state energies in the last column were obtained  
from the solution of the Schr\"odinger equation in each given  
potential.
    Notice  
that the breakup cross sections for potentials with negative scattering  
lengths
are much larger than those with positive scattering lengths.
This effect is mainly due to the influence of the phase $\nu$.

The effective phase in Eq.(\ref{soff}) is $\bar \delta= \delta+ \nu$ and there are interference effects which are especially important for an s-state final state. When $k$ is small $\delta\approx -k a_s$ and it depends on the sign of the scattering length, while $\nu \approx k(b_c+1/\gamma)$ and it is always positive. The part of the cross section Eq.(\ref{cross}) with the probability Eq.(\ref{8}) for $l_f=0$ and small $k$, depends on the relative sign of a$_s$ and $\nu$  
 as
\begin{equation}{d\sigma\over {d{{\varepsilon_f}} }} \propto |\delta+\nu|^2 \approx |-ka_s+\nu|^2.\label{delnu}\end{equation}

When a$_s<0$ the cross section is increased relative to the value at $\nu=0$, while for a$_s>0$ it is reduced. This effect is seen clearly in Fig.\ref{fig3a}. These interference effects can also explain why the cross section for a$_s$=4.8 fm is larger than for a$_s$=7.1 fm.  Also, the decrease in the positive values of a$_s$ shown in Table \ref{r} corresponds to an increase in the depth of the potential $\delta$V. Fig. \ref{fig3a} shows that the cross section increases for the smallest values of a$_s$. As the attraction becomes ever stronger the scattering length changes sign and the cross section becomes larger. This effect is due to the influence of a continuum s-state coming close to threshold. When the final potential has a very weakly bound 2s-state with a$_s$=22.4 fm one  sees  a  
very narrow peak close to threshold (thick solid line) while for a$_s$= 4.5 fm, corresponding to a more strongly bound state, no peak at  
all, rather a kind of flat bump (double-dotdashed line).

\begin{figure}[h]
\vskip20pt
\center \scalebox{0.4}
{\includegraphics{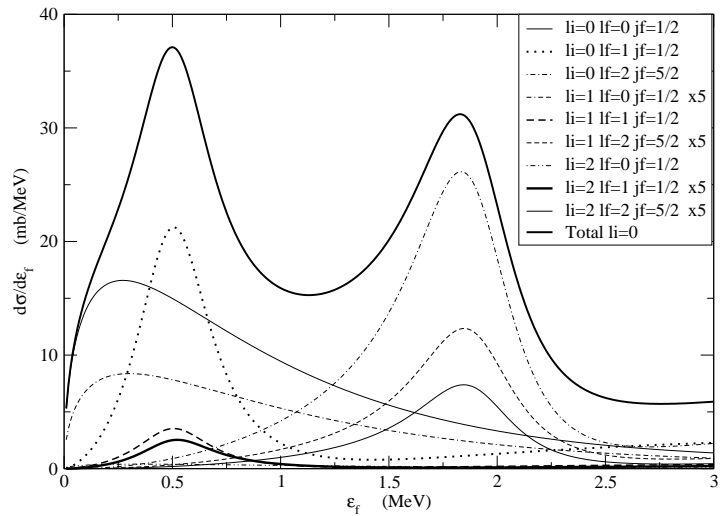}}
\caption {\footnotesize  Results obtained including the s, p and d  states. Each curve corresponds to just one transition as indicated. The solid curve is the sum of all transitions from the s-bound state. To make them visible some curves have been multiplied by   a factor of five as indicated in the legenda.}
\label{fig5}
\end{figure}

The  
calculations presented in Fig. 5 of Ref.\cite{bbv},  did not include  
the extra phase $\nu$ because the final state interaction with the core of  
origin  was neglected  while the  final state interaction with respect  
to the other nucleus was included. In that case the neutron behaved as a free  
particle in the scattering on the other nucleus. Breakup cross sections  
depended quite strongly on the magnitude of the
scattering length but had a weak dependence on its sign. The reaction  
mechanism discussed in this paper is instead an inelastic-type of  
excitation in which the final state interaction is with the  core of  
the projectile and therefore the present formalism shows that the S-matrix in  
Eq.(\ref{5tris}) as well as in Eq.(\ref{geo}) is effectively off-the-energy-shell.
 In the s $\to $ s case we show also in Fig.\ref{fig3b}, a comparison between the results
just discussed and those obtained using the optical model phase shifts in Eq.(\ref{geo}) and whose absolute values have been normalized to our peaks. 
As anticipated in Sec. 2.3 and 2.4, the  curves from the time dependent model show a faster decrease towards high energies. This is  because the form factors in Eq.(\ref{8}) decrease rapidly  and they have large values only for impact parameters close to the strong absorption radius. 
\begin{figure}[h]
\vskip20pt
\center \scalebox{0.4}
{\includegraphics{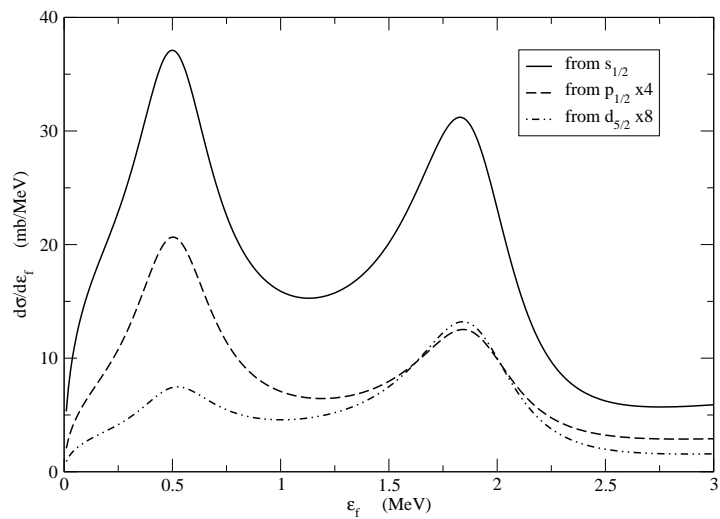}}
\caption{\footnotesize Check of the dependence from the initial state angular momentum. Full  curve: sum of transitions from s-initial state. Dashed and dotdashed lines:  sum of transitions from p and d-initial states respectively. To make them visible these curves have been multiplied by   a factor of four and eight respectively. }
\label{fig6b}
\end{figure}

Since Eq.(\ref{geo}) is quite simple to implement in an fit to experimental spectra, we have also studied its sensitivity to the choice of the phase shift. Fig.\ref{fig4} shows, for two scattering lengths, results obtained using  optical model phase shifts, the second order effective range approximation Eq.(\ref{er}) with values from Table \ref{r}, and the first order phase shift
$\delta \simeq -a_s k$, as indicated. As expected, the latter approximation is reliable only for extremely small values of the final energy. The second order, effective range parametrization works much better, in particular as the scattering length
increases.  

\subsection{ p and d-resonances }

Fig. \ref{fig5} shows results obtained considering  three different possibilities  
for the initial state: the s, p and d orbitals.  For each initial state a unit  
spectroscopic factor is assumed.

 \begin{figure}[h]
 \vskip 20pt
\center \scalebox{0.4}
{\includegraphics{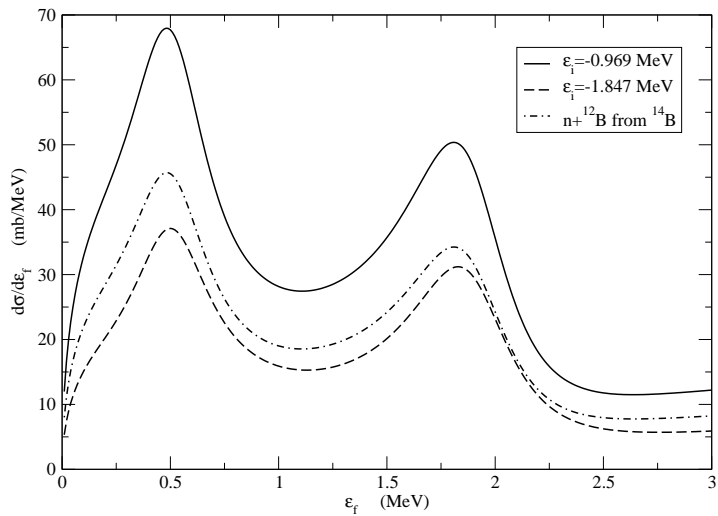}}
\caption{\footnotesize Check of the dependence from the initial binding  
energy of the sum of transitions from s-initial state. Full curve: $\varepsilon_i$= -0.97 MeV as in $^{14}$B; dashed curve: $\varepsilon_i$=-1.85 MeV as  in  
  $^{14}$Be \cite{nucbase}. Dotdashed line: sum of transitions from s, p and d-initial states including spectroscopic factors of 0.66, 0.04. 0.30 respectively as in $^{14}$B \cite{msu} with $\varepsilon_i$= -0.97 MeV.}
\label{fig6a}
\end{figure} 

We show the results of the transition bound to unbound from each initial  
state to each possible  unbound state as indicated. Available experimental spectra \cite{Belo98,Kor95} show that the next group of resonances is located around 4-5 MeV. For  this reason  higher partial  waves have not been included. We have checked that if the    d$_{3/2}$ and p$_{3/2}$ states are calculated in the same potentials as those of the  d$_{5/2}$ and p$_{1/2}$, they would give  a noticeable,  non resonant, contribution only for a transition from the s initial state. This contribution, not shown in the figures, would constitute a smooth background. The overall  effect would be an increase the  cross section value of about 10$\%$ around the 0.5 MeV peak  and of about 25$\%$  around 2 MeV in  Fig.\ref{fig5}. 
The spectrum for Coulomb breakup has also been calculated and found negligible, compared to the other transitions, for the separation energies in  $^{14}$Be and $^{14}$B. Therefore it is not included in the figures.

 The p and the d-resonance peaks are clearly seen  
because of the effect of the  angular momentum enhancement factor in Eq.(\ref{54}). As indicated some transition strengths have been multiplied by a factor five to make them visible. 
It is clear from this figure that the dominant components
in the neutron knockout spectrum from $^{14}$Be and $^{14}$B come from the s-wave component in the ground states of those nuclei. 
Therefore the full curve is the sum of the contributions from the initial s-state alone  with
 unit spectroscopic factor. There can be  large peaks due to  transitions to the 
p$_{{\footnotesize {1/2}}}$ and  d$_{5/2}$ final state, provided they are centered around  the energies we have choosen.
 The results of  Fig. \ref{fig6b} are shown to check  the dependence of the transition probability  on  the initial state angular momentum. The full  curve is the  sum of transitions from s-initial state. Dashed and dotdashed lines are the sum of transitions from p and d-initial states respectively. Since the transitions from p and d orbitals are negligible,  then 
components in the wave functions of $^{14}$Be or $^{14}$B with a neutron  
in such angular momentum states
will not play much role in the reaction. Thus it is unlikely that any  
difference in the neutron
  breakup spectrum due to different mixtures of these configurations in  
the two parent nuclei will be seen.

To clarify further the latter point, the sum of all transitions from the s-bound state,   is   
shown again in Fig. \ref{fig6a} for two initial binding energies (solid and dashed lines as indicated). Very small changes in the  
initial binding energy do not give
appreciable differences in the final continuum spectra, in particular  
in the positions of the peaks. They however give differences in the  
absolute cross section value and in the relative height of the s- and  
d-state peak. The dotdashed line is the result obtained using the neutron initial binding energy in $^{14}$B, which is -0.969 MeV, and summing transitions from s, p and d-initial states including experimental spectroscopic factors \cite{msu} of 0.66, 0.04. 0.30 respectively.  
According to the simple model for two nucleon breakup presented in Sec. 3,
the presence of a 1p$_{3/2}$ proton in the initial state would give the
same contribution to the breakup of each neutron component and therefore the shape of the spectra should not be modified. 
On the other hand in the case of $^{14}$Be the two neutrons are in the same  state for each component of the initial wave function. Therefore since the p and d wave functions have less pronounced tails, the last integral of Eq.(\ref{cross123}), which gives the stripping probability of one of the two neutrons, will naturally diminish the absolute value of the p and d resonances peaks with respect to the final s-state peak. 

We notice also that there is a shift between  the peaks of the energy spectra and the resonance energies obtained from the phase shifts and given in Table 4. The shift is due to a combined effect of  the 1/k factor in Eq.(\ref{8}), of the matching between initial separation energy, peak energy of the resonance and relative beam energy per nucleon and only in a small measure to the presence of the extra phase $\nu$. 
The matching effect is manly contained in the slope of the form factor, which to first order is equal to the decay length of the initial state wave function (cf. Eq.(\ref{k})). In the $^{11}$Be case it is less noticeable because the initial separation energy  is  very small. The resonance energy was 1.27 MeV while the peak is at 1.25 MeV.  For $^{13}$Be the  resonance energies are 0.67 MeV (p-state) and 2 MeV (d-state) while the peaks are at 0.5 MeV and 1.8 MeV respectively.  One notices also that the shift increases increasing the resonance energy.

Therefore two differences might be expected  in the experimental energy spectra of  $^{13}$Be when produced by  $^{14}$Be rather than  $^{14}$B projectile fragmentation: a first peak
at energy smaller than 0.5 MeV due to the s-continuum state. The s-continuum peak below 0.5MeV, as given by our calculations, can be seen better in Fig.8 and it is due to the s to s transition. If the s initial component would contribute in reality more than the p and d components (we have taken them with equal weights) such a peak could be more evident in the data.
 Then two more diffuse bumps  if the projectile is 
$^{14}$Be. Two well definite peaks of not too different strength, one centered at 0.5 MeV due to the p$_{1/2}$ resonance and another around 2 MeV due to the d$_{5/2}$ resonance if the projectile is $^{14}$B. Such an hypothesis agrees also with the conclusion of Ref.\cite{Thoen00} of a  s-state very close to threshold. The three different experiments \cite{jl,Thoen00,simo} would therefore be complementary and allow to determine the characteristics of $^{13}$Be low energy continuum.
\begin{figure}[h]
 \vskip 25pt
\center \scalebox{0.4}
{\includegraphics{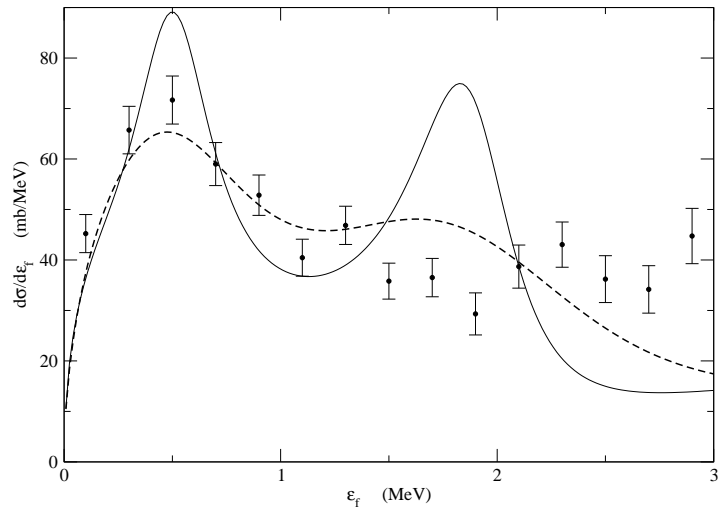}}
\caption {\footnotesize  Sum  
of all transitions from the s initial state with $\varepsilon_i$=-1.85 MeV (solid line). Experimental points from H. Simon et al.  \cite{lc} for the reaction $^{14}$Be+$^{12}$C $\to$ n+$^{12}$Be+X at 250 A.MeV. Dashed line is the folding of the calculated spectrum with the experimental resolution curve.}
\label{fig12}
\end{figure}

To give another example of a possible comparison with available  data, we show in Fig. \ref{fig12} the experimental points from H. Simon et al. \cite{lc} for the reaction $^{14}$Be+$^{12}$C $\to$ n+$^{12}$Be+X at 250 A.MeV. The normalization factor of the data  to mb/MeV is 0.843. The solid line gives the sum  
of all transitions from the s initial state with $\varepsilon_f$=-1.85 MeV (solid line), as in previous figures, renormalized with a factor 2.4.  The dashed line is the folding of the calculated spectrum with the experimental resolution curve. Therefore the calculation underestimate the absolute experimental cross section by a factor of 2.   In view of the incertitude in the strength of our n-target $\delta$-potential and on the initial state spectroscopic factor which has been taken as unit, we can consider   our absolute cross sections  quite  reasonable. 

\begin{figure}[h]
 \vskip 20pt
\center \scalebox{0.4}
{\includegraphics{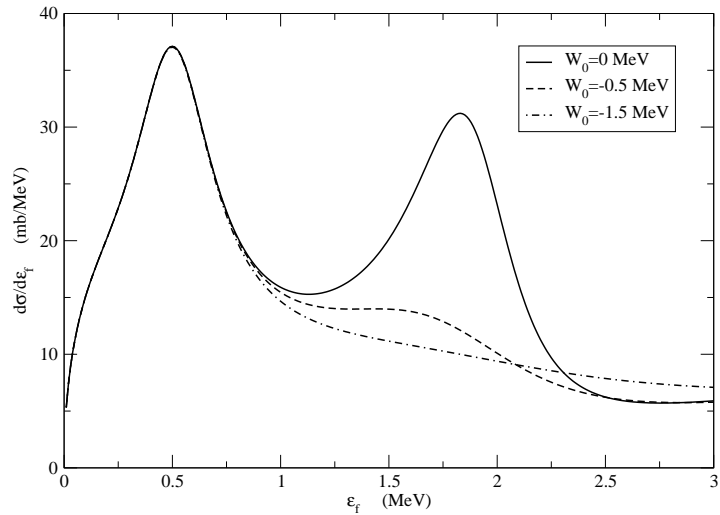}}
\caption {\footnotesize  Sum  
of all transitions from the s initial state with $\varepsilon_f$=-1.85 MeV including  
core excitation via an imaginary part of the optical potential for the  d-resonance only.}
\label{fig7}
\end{figure}

Finally we wanted to address the issue of possible core excitation  
effects which in Ref.\cite{ta} have been shown to be of fundamental
importance for reproducing simultaneously the $^{13}$Be and $^{14}$Be  
characteristics. Those effects can be modeled    in the present
approach by considering a small imaginary part in the neutron-core  
final optical potential (cf. Eq.(\ref{op})). This is a standard procedure for continuum states where most often the potential is also energy dependent \cite{mahux}.
 A surface potential of  
Woods-Saxon derivative
form has been taken with strengths W$_0$ equal to -0.5, -1.0 and -1.5 MeV for the d-state only. The  
results are shown in Fig. \ref{fig7}. The effect of the imaginary potential
is to wash out the  
d-resonance peak. Several structure models predict indeed  the d$_{5/2}$ resonance coupled to  an excited $^{12}$Be core. We have found the same smoothing off effect if the s-state is calculated including an imaginary potential. It seems therefore  
that the spectrum of unbound
nuclei could partially reflect the structure of the bound parent nucleus and that  
reaction mechanism models used to extract structure information
should carefully include the effects discussed above. The model  
presented here seems to be quite promising in this respect.

\section{Conclusions and outlook}

In this paper we have presented a model to study one neutron  
excitations from a bound initial state to an unbound resonant state in  
the
neutron-core low energy continuum. This is the process by which unbound  
nuclei are created and studied via projectile fragmentation
experiments \cite{jl}-\cite{6he,lc}.

The model is based on a time dependent perturbation theory amplitude  
and the final state is described by an optical model S-matrix.
It can be considered an evolution with respects to sudden and/or  
R-matrix theory models. The advantages are that the model can be
applied to fragmentation from deeply bound states and to resonant and non resonant,  
large energy, continuum final states. Also core excitation
effects can be modeled by an imaginary part of the neutron-core  
optical potential.

 A related approach has been developed some time
ago and applied to the treatment of transfer to the continuum in which,  
following the interaction between two passing-by nuclei, a
neutron from one of them goes
from a bound to an unbound state with final state interaction with the  
other nucleus. Comparison of the present formalism to the
transfer to the continuum model shows that in principle projectile  
fragmentation does not reflect directly the properties of the
neutron-core resonances because the reaction mechanism induces an extra  
phase  with respect to the free particle neutron-core
phase shift. It means that the measurements would probe an  
off-the-energy-shell S-Matrix. The distortion effects seem however  
small and negligible for the cases discussed in this work.
 
 One neutron breakup can be studied in this way  
but also one step of two neutron breakup of a borromean nucleus. 
In  
this paper we have presented some applications to both cases  to
study the properties of  $^{11}$Be continuum and of  $^{13}$Be. Our results are in agreement with  
the conclusions of Ref.\cite{fuku,cap} for $^{11}$Be. Due to the structure inputs we use,  in particular the position of the p$_{1/2}$ resonance, the
$^{13}$Be continuum spectrum obtained from fragmentation of   $^{14}$B or  $^{14}$Be
shows essentially the effect of the continuum p and
d-resonances. The s-state  although present in the  
calculations almost disappears inside
the tail of the p-state but it  would still determine  the ground state spin and parity of $^{13}$Be. 
Obviously we cannot be conclusive on the structure of $^{13}$Be because at the moment we have not attempted to fit  experimental data but simply to develop a good reaction model. Furthermore our structure inputs, although reasonable, are extremely simple compared to the complexity of the nucleus under study. However preliminary comparisons seem to indicate the reliability of our model.

 We have also shown that the excitation energy spectra of an  
unbound nucleus might reflect the structure of the parent nucleus from whose fragmentation they are obtained.  In particular, in the case of $^{14}$Be fragmentation,  the initial state spectroscopic factors are not known experimentally, and the information from  structure calculations indicate an important configuration mixing with components coupled to an excited $^{12}$Be core. Thus the analysis of such spectra is expected to be even more complicated. 

However from the  point of view of reaction theory, the results obtained  here
seem
promising and we hope  to use such a procedure to implement a fit of experimental data on unbound nuclei. At the same time we intend  to construct an accurate, second order, fully time dependent
theory of two neutron breakup, incorporating properly the time ordering between the two neutrons. In this way we hope to clarify the question of sequential versus simultaneous mechanism implicit in the formation of  neutron-core resonance states  
in reactions like
$^{11}{\rm Li+X} \to ^{10}$Li  $\to ^{9}$Li+2n
\cite{jl} or
$^{14}{\rm Be+X} \to ^{13}$Be$^*$+n + X$\to ^{12}$Be+2n + X, or
$^{14}{\rm B+X} \to ^{13}$ Be$^*$+p + X$\to ^{12}$Be+n +p + X.

{\bf Acknowledgments}

\noindent We wish to thank Bj\"orn Jonson and his collaborators,  in particular Leonid Chulkov and Haik Simon, for communicating their results previous to publication, for
folding our calculations with their experimental resolution and for an enlightening correspondence. Further discussions with  T. Aumann, G. Bertsch, U. Datta-Pramanik, K. Jones, N. Orr, and S. Shimoura, are gratefully  acknowledged.
This work was inspired by discussions with  Takashi Nakamura and the late P. Gregers Hansen  
at the {\it Spectroscopic Factors} workshop, ECT*, Trento, 2004. 
\appendix\section{Modifications to the $ {\mathbf \delta}$-interaction}
\begin{figure}[h!t]\caption{Graph of variables used in the calculation of appendix A.}
 \vskip 20pt
   \begin{center}      \scalebox{0.5}{
              \includegraphics{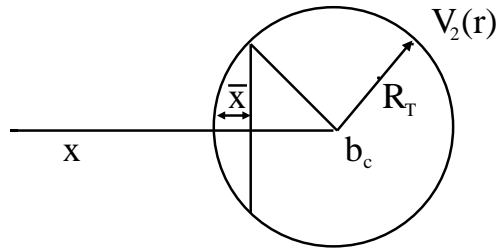}}                    
  \label{vargraf}   \end{center}  \end{figure}

The purpose of this appendix is to justify the use of a $\delta$-interaction as an approximation for the finite range n-target interaction and to derive Eq.(\ref{ap1}).
We then calculate 
\begin{equation}
J= \int_{-\infty}^{\infty}dxdydzdz^{\prime} ~ {e^ {-(\gamma - ik) r}\over r^2}e^{-iq(z-z^{\prime})}V_2(x-b_c,y,z^{\prime}).\label{ap2}
\end{equation}
If $\gamma$ is large the integral is concentrated near the surface of V$_2$(r). To simplify the discussion put $q=0$.
Also, as in Sec. 2.3 put
\begin{eqnarray}
 e^ {-(\gamma - ik) r}& \approx & e^ {-\alpha x} e^{-\alpha(y^2+z^2)/2x} \nonumber \\ &\approx& e^ { -\alpha x} e^{-\alpha z^2/2R_0},\label{ap4}
\end{eqnarray}
where $\alpha=\gamma-ik$ and we used the fact that the gaussian term $e^{-\alpha(y^2+z^2)/2x}$ gives the largest contribution at $ {\rm R}_0={\rm b}_c- {\rm R}_T$ which is the position of the surface of V$_2$(r). 
To simplify further the calculation we have  neglected the y-dependence in Eq.(\ref{ap4}) but kept the z-dependence so that the integral J will converge. 
  Indicate ${\bar x}=x- {\rm R}_0=x-({\rm b}_c- {\rm R}_T)$.
 Take V$_2$(r) to be a square well potential of depth V$_0$ and radius
 R$_T$. 
 Then \begin{equation}
 \int_{-\infty}^{\infty}~ V_2(x-b_c,y,z^{\prime})~dy dz^{\prime} =\pi V_0 \beta^2=2\pi V_0  R_T \bar x \label{ap5}
 \end{equation} where the upper limit of the two dimensional integral is given by $\beta^2+(\bar x- {\rm R}_T)^2= {\rm R}_T^2$ and $\beta^2\approx 2\bar x {\rm R}_T$.
 
So that
  \begin{eqnarray} J&=& 2\pi V_0  R_T\over b_c^2}{ \int dx e^ {-\alpha x} \bar x \int dz e^{-{\alpha z^2\over 2R_0}} \nonumber \\ &=&{V_0\over b_c^2}\sqrt {{2\pi(b_c-R_T)\over {\gamma -ik}}}e^ {-\alpha (b_c-R_T)} 2\pi   R_T\int_{b_c-R_T}^{\infty} d\bar x e^ {-\alpha \bar x} \bar x \nonumber \\ &=&{V_0 \over b_c^2}\sqrt {{2\pi(b_c-R_T)\over {\gamma -ik}}}2\pi   {R_T\over \alpha^2}e^ {-{\alpha }(b_c-R_T)}
 \label{ap6}
\end{eqnarray}
 The ratio of the integral J to the integral I of Eq.(\ref {14}) is 
 
  \begin{equation}{J\over I}=\left({b_c-R_T\over b_c }\right)^{{1\over 2}}e^{(\gamma-ik) R_T}{3\over 2}{v_2\over R_T^2(\gamma-ik)^2},
 \label{ap8} \end{equation}
Where we imposed that the strength of the $\delta$-interaction be equal to  the volume integral of the square well potential ${v}_2={4\over 3} \pi V_0  R_T^3$. Thus to represent the effect of a finite range potential by a $\delta$-interaction when $\gamma {\rm R}_T>>1$, replace
\begin {itemize}
\item {(1) ${\rm b}_c\to {\rm b}_c-{\rm R}_T$ i.e. the interaction is at the surface of the target.}

\item { (2) Multiply the strength of the interaction by $ {3\over 2}{1\over{\rm R}_T^2(\gamma-ik)^2}$. }
 
\end{itemize}
This factor is less than one. The change (1) increases the breakup integral, the factor (2) decreases it.

  \section{Including spin}

To include spin variables in the initial and final states
is mainly an angular momentum coupling problem.
The angle-spin wave function of the initial and final states are
\beq
\Psi_i(j_i, n_i, l_i, \theta, \phi) = \sum_{m_i \sigma_i}
\langle j_i n_i|l_im_i{  \frac {1}{ 2}} \sigma_i\rangle Y_{l_i m_i}(\theta, \phi) \chi_{\sigma_i}(\rho)
\eeq
\beq
\Psi_f(j_f, n_f, l_f, \theta, \phi) = \sum_{m_f \sigma_f}
\langle j_f n_f|l_fm_f {\footnotesize {1\over 2}} \sigma_f\rangle Y_{l_f m_f}(\theta, \phi) 
\chi_{\sigma_f}(\rho).
\eeq
We choose  the quantization axis along the y-direction, such that $\phi=0$.
Then after integration over $\rho$, the angle spin part of the overlap Eq.(\ref{1bis})  is: 
\bea
 {\cal D}(j_fn_f, j_i n_i) &=&\sum_{m_f m_i \sigma}
\langle j_f n_f|l_fm_f{\footnotesize {1\over  2}} \sigma\rangle \langle j_i n_i|l_im_i {\footnotesize {1\over 2}} \sigma\rangle 
Y^*_{l_f m_f}(\theta, 0) Y_{l_i m_i}(\theta, 0) \nonumber \eea \bea =\sum_{m_f m_i \sigma}(-1)^{m_f}
\langle j_f n_f|l_f -m_f {\footnotesize {1\over 2}} \sigma\rangle \langle j_i n_i|l_im_i {\tiny 1\over 2} \sigma\rangle 
Y_{l_f m_f}(\theta, 0) Y_{l_i m_i}(\theta, 0),
\eea
where we have put $Y^*_{lm} = (-1)^mY_{l-m}$.  Next we use the relation for coupling two spherical harmonics of the same argument and introduce the notation  $\hat i=\sqrt{2i+1}$
\beq
Y_{l_i m_i}(\theta,0)Y_{l_f m_f}(\theta,0)= \sum_{LM} \langle LM| l_im_i l_f m_f \rangle 
\langle L0| l_i 0 l_f 0 \rangle 
{\frac {{\hat l_i }{\hat l_f }}{\sqrt {4\pi} {\hat L}}} Y_{LM}(\theta, 0).
\eeq
Substituting into the relation for ${\cal D}(j_fn_f, j_i n_i)$ there is a sum of products of three Clebsch-Gordan coefficients which reduces to a product of a Clebsch-Gordan coefficient and a 6-j symbol. 
Collecting together the terms evaluated above  we get:
\[
{\cal D}(j_fn_f, j_i n_i)= (-)^{f}\sum_{LM} \langle LM| j_in_i j_f -n_f \rangle 
\langle L0| l_i 0 l_f 0 \rangle 
\frac{{\hat l_i}{\hat l_f}{\hat j_i}{\hat j_f}}{\sqrt {4\pi} {\hat L}}Y_{LM}(\theta, 0)
\left\{\begin{array}{ccc}
l_i & l_f & L \\
j_f &  j_i & {\footnotesize {1\over 2}}
\end{array}
\right\}.
\]
With the phase $f=n_f+l_f-j_f$

In this scheme the   integral Eq.({\ref{8bis}) 
\begin {equation}  I_{l_i m_i,l_f m_f} =
\int_{-\infty}^{\infty} dze^{iqz}i^{l_i}
 \gamma h^{(1)}_{l_i}(i\gamma  r)Y_{{l_i},{m_i}}(\theta,0)  
k{i\over 2}h^{(-)}_{l_f}(kr)Y_{{l_f},{m_f}}(\theta,0)
\label{a8bis} 
\end{equation} 
is substituted by a new integral $I_{LM}$ defined as
\begin {equation}  
I_{LM} =\int_{-\infty}^{\infty} dze^{iqz}i^{l_i}
 \gamma h^{(1)}_{l_i}(i\gamma  r)
  k{i\over 2}h^{(-)}_{l_f}(kr)Y_{{L,M}}(\theta,0).
\label{a8tris} \end{equation} 
Summing over $n_f$ and averaging over $n_i$ and using the orthogonality of the $\langle LM| j_in_i j_f n_f \rangle 
$ coefficients while calculating $|{A}(j_fn_f, j_i n_i)|^2$  as in Eq.(\ref{g}) we find that
 Eq.(\ref{8}) can be replaced by:

\beq
{dP_{in}\over d\varepsilon_f}={2\over \pi}{v_2^2\over \hbar^2  
v^2}{C_i^2 }{m\over\hbar^2k}\sum_{LM} C(l_i,j_i,l_f,j_f;L)|\langle L0| l_i 0 l_f 0 \rangle |^2
 |1-\bar S_{LM}|^2 |I_{LM}|^2, \label{54}
\eeq
where
\beq
C(l_i,j_i,l_f,j_f;L)= \frac{(2j_f+1)(2l_i+1)(2l_f +1)}{4\pi (2L + 1)} 
\left\{\begin{array}{ccc}
l_i & l_f & L \\
j_f &  j_i & {\footnotesize {1\over 2}}
\end{array}
\right\}^2.
\eeq

A sum rule for $6-j$ symbols gives
\beq
\sum_{j_f}C(l_ij_i,l_f,j_f;L) = \frac{(2l_f +1)}{4\pi (2L + 1)}. 
\label{amc2}
\eeq

On the other hand, if in Eq.(\ref{8}) and (\ref{8bis}) or (\ref {a8bis}) we use
the angular momentum coupling formula for spherical harmonics  
\beq
\sum_{m_i m_f} \langle LM| l_im_i l_f m_f \rangle 
Y_{l_i m_i}(\theta,0)Y_{l_f m_f}(\theta,0)= \langle L0| l_i 0 l_f 0 \rangle 
\frac {\hat {l_i} \hat {l_f}} {\sqrt {4\pi} \hat {L}}Y_{LM}(\theta, 0),
\label{amc}
\eeq
then the relation between $I_{m_i m_f}$ and $I_{LM}$ is 
\beq
I_{l_i m_i,l_f m_f}=\sum_{LM} (-1)^{m_f}\langle LM| l_f -m_f l_i m_i\rangle 
\langle L0| l_i 0 l_f 0 \rangle 
\frac {\hat {l_i} \hat {l_f}} {\sqrt {4\pi} \hat {L}} I_{LM}
\label{Ililf}
\eeq
and Eq.(\ref{8}) is replaced by
\begin{equation}
{dP_{in}\over d\varepsilon_f}={2\over \pi}{v_2^2\over \hbar^2  
v^2}{C_i^2 }{m\over\hbar^2k}\sum_{LM}
\frac{(2l_f +1)}{4\pi (2L + 1)} 
|\langle L0| l_i 0 l_f 0 \rangle |^2
 |1-\bar S_{LM}|^2 |I_{LM}|^2,
\label{a8tris}
\end{equation}
which could also be obtained using Eq.(\ref{amc2}) in (\ref{54}).

\end{document}